%% file: for_arxiv.tex
\documentclass[a4paper]{article}
\usepackage{times}
\usepackage{graphicx}
\usepackage[margin=3cm]{geometry}
\usepackage{amsmath}
\usepackage{amssymb}
\usepackage{bbm} 
\usepackage{color}
\usepackage[utf8]{inputenc}
\usepackage{hyperref}
\usepackage[superscript]{cite}
\usepackage{authblk}

\title{
Human learning for molecular simulations:\\ the Collective Variables Dashboard in VMD
}
\author[,1,2]{J\'er\^ome H\'enin\thanks{Corresponding author: \texttt{henin@ibpc.fr}}}
\author[3]{Laura J. S. Lopes}
\author[4]{Giacomo Fiorin}
 
\affil[1]{Laboratoire de Biochimie Théorique UPR 9080, CNRS, Université de Paris, 75005 Paris, France}
\affil[2]{Institut de Biologie Physico-Chimique--Fondation Edmond de Rothschild, PSL Research University, 75005 Paris, France}
\affil[3]{Theoretical Division T-1, Los Alamos National Laboratory, Los Alamos, New Mexico 87545, USA}
\affil[4]{Theoretical
 Molecular Biophysics Laboratory, National Heart,
Lung and Blood Institute, National Institutes of Health, 10 Center Drive, Bethesda,
MD 20814}
 
\begin{document}
\maketitle

\begin{abstract}
\normalsize

The Collective Variables Dashboard is a software tool for real-time, seamless exploration of molecular structures and trajectories in a customizable space of collective variables.
The Dashboard arises from the integration of the Collective Variables Module with the visualization software VMD, augmented with a fully discoverable graphical interface offering interactive workflows for the design and analysis of collective variables.
Typical use cases include a priori design of collective variables for enhanced sampling and free energy simulations
as well as post-mortem analysis of any type of simulation or collection of structures in a collective variable space.
A combination of those cases commonly occurs when preliminary simulations, biased or unbiased, reveal that an optimized set of collective variables is necessary to improve sampling in further simulations.
Then the Dashboard provides an efficient way to intuitively explore the space of likely collective variables,
validate them on existing data, and use the resulting collective variable definitions directly in further biased simulations using the Collective Variables Module.
We illustrate the use of the Dashboard on two applications: discovering coordinates to describe ligand unbinding from a protein binding site, and designing volume-based variables to bias the hydration of a transmembrane pore.
\end{abstract}

\bigskip
{\bfseries Keywords: molecular dynamics, collective variables, visualization, analysis}

\section{Introduction}

Research using biomolecular modeling and simulations nearly always resorts to geometric descriptors for analyzing molecular trajectories, or structures and their differences.
Such descriptors are low-dimension functions of atomic coordinates, and are often referred to as \emph{collective variables} (abbreviated to colvars).
Collective variables can be used either for post-mortem analysis of sets of structures and simulated trajectories, or to modify simulations on the fly in order to restrain, enhance, or otherwise bias the motion of chosen degrees of freedom.
When preparing a biased simulation, these variables are chosen based on previous knowledge.
Constructing such a reduced space of relevant coordinates constitutes a dimensionality reduction process.
This step may be performed as an automated, data-driven approach, where dimensionality reduction algorithms determine optimal collective variables for a given problem.\cite{Sittel2018, Tribello2019}
Still, these approaches are a topic of active research, and have not reached the stage where they can be applied as routine, black-box tools.
Furthermore, an intuitive interpretation of the reduced representation is always desirable eventually.
Some dimensionality reduction methods are explicitly aimed at constructing interpretable coordinates.\cite{McGibbon2017} 
In most current applications, the dimension-reduction process is knowledge-based rather than data-driven: the practitioner's intuition, guided by available information, is the driving force.
Choosing and refining the relevant collective variables requires a large expense of human time and effort.
Thus, any workflow using collective variables sooner or later requires an intuitive connection between the space of atomic coordinates and the space of collective variables.
These variables are typically analyzed and visualized using combinations of more or less ad hoc analysis software and generic plotting tools.

The new tool we publish here provides an immediate, intuitive connection between existing molecular data and a very broad space of possible collective variables, resulting in more robust reduced descriptions with reduced researcher time and effort.
It provides a concrete, visual sense of the abstract mathematical objects that are collective variables.
In our hands, this has proven just as useful to scientists in training as to experienced practitioners.

A fundamental everyday tool of molecular biophysicists is molecular graphics software, such as VMD.\cite{Humphrey1996}
In addition to graphics, VMD already provides some ways to access geometric measurements on static structures or along trajectories.
The simplest form consists in picking pairs, triplets and quadruplets of atoms with the mouse to label distances, angles and dihedral angles respectively.
Analysis plugins offer more variables that can be analyzed, notably root-mean-square deviations (RMSD), H-bonds and salt bridges, and more collective properties such as contact maps, radial distribution functions, or projections along normal modes.
These represent only a limited subset of the variables that are used for biasing or analyzing simulations.

The Collective Variables Module (Colvars)\cite{Fiorin2013} is a software library integrated with the standard releases of major molecular simulation and analysis software packages, including LAMMPS\cite{Thompson2021}, NAMD\cite{Phillips2020} and VMD\cite{Humphrey1996}.
Additionally, patched releases of GROMACS\cite{Abraham2015} have also become available recently.
Colvars can be used to modify the simulated dynamics according to various types of biasing algorithms acting on functions of the atomic Cartesian coordinates; we refer to those functions as collective variables, or \emph{colvars} in short.
The Colvars Module implements adaptively biased, free energy simulations such as variants of the Adaptive Biasing Force (ABF)\cite{Darve2001,Henin2010a,Lesage2017} and metadynamics\cite{Laio2002,Barducci2008}.
The complete space of collective variables can be used for adaptive free energy calculations using eABF\cite{Lesage2017}, which lifts some technical requirements of ABF\cite{Darve2001, Comer2015}.
Within VMD, the Collective Variables Module functions as an analysis tool, allowing for computation and analysis of colvars through a scripting interface based on the Tcl language used by VMD's command line interface.

\begin{figure}[!ht]
\centering
 \includegraphics[width=7cm]{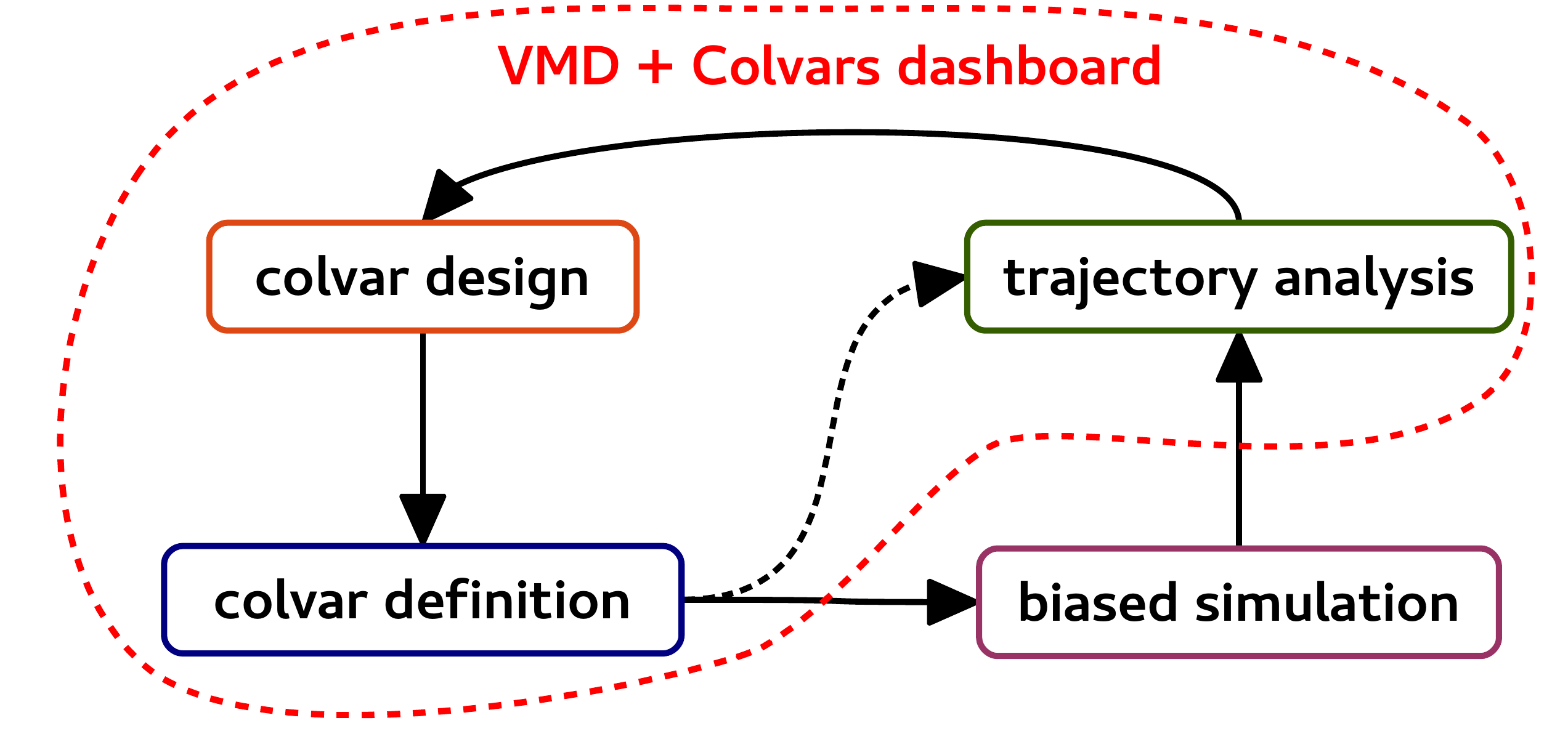}
 \caption{A generic collective variables modeling workflow. Several steps are unified and simplified using the Dashboard, and transfer between analysis and simulation is streamlined thanks to the compatible Colvars Module interface.}
\label{fig:workflow}
\end{figure}

The workflow typically begins with exploratory, unbiased simulations.
The resulting trajectories can inform the design of biased simulations aimed at more fully exploring regions of interest and sampling those metastable degrees of freedom that are deemed relevant by the scientist.
Information gleaned from biased simulations may in turn suggest an improved space of collective variables, resulting in the iterative workflow illustrated by the solid arrows in Figure~\ref{fig:workflow}.

This process is limited by two types of obstacles.
The fundamental obstacle is the difficulty of defining and discovering the most relevant degrees of freedom in high-dimensional systems; this may require iterations on any existing structural or simulation data.
Technical obstacles arise in the various steps necessary to design, implement and calculate collective variables on-the-fly or when post-processing simulations.
The Collective Variables Module (\textit{Colvars}) is a broad-purpose toolbox, and writing correct input files making use of all its features requires some time and care.
Analyzing trajectories may involve running several pieces of specialized software in succession, either manually, or with the help of in-house scripts.
The Colvars Dashboard is a software helper designed to accelerate this process going from data to information, to knowledge.
Even In the case of a purely data-driven dimensionality reduction such as PCA, dihedral PCA\cite{Altis2007}, FCA\cite{Lange2008} or TICA,\cite{Molgedey1994, Perez-Hernandez2013, Schwantes2013} the Colvars Dashboard allows for inspecting the data in the reduced space, to build \textit{a posteriori} a physical intuition about the reduced space.

There are currently graphical interfaces for VMD that interact with the collective variable software PLUMED~\cite{Tribello2014}: Plumed-GUI~\cite{Giorgino2014} as a writing aid, and METAGUI~\cite{Biarnes2012,Giorgino2017} for clustering and free energy analysis. Both plugins invoke PLUMED as an external executable, passing data from VMD's memory as temporary files; in addition, PLUMED-GUI offers configuration templates and an interface for plotting variables using the Colvars Module.
However, the present tool provides a much broader access to the functionality of the Colvars Module. 
In particular, it exploits the binary-level integration of Colvars with VMD to provide a fully interactive workflow where variables are updated in real-time and the molecular and collective variable visualizations are tightly linked.

Use cases for the Colvars Dashboard include:
\begin{itemize}
 \item Defining colvars \emph{a priori} based on a single known structure;
 \item Exploring different colvars on a set of known structures;
 \item Analyzing simulated trajectories in colvar space;
 \item Refining colvars based on simulated trajectories.
\end{itemize}

Below we describe the design and implementation of the Dashboard, explain its uses through two types of typical tasks (visualizing a given set of variables, and designing new variables). We then illustrate its functionality on two real-world scientific applications: exploring protein-ligand binding and unbinding, and designing and visualizing variables based on volumetric maps.

\section{Design and implementation}

The functional, user-visible design of the Dashboard is summarized in Table~\ref{tab:tasks}.

\begin{table}[!ht]
\begin{tabular}{l|l}
 \textbf{Task} & \textbf{Technical solution} \\
 \hline
 explore previously defined colvars & load config file\\
 use colvars from Dashboard in future simulations & save config file\\
 \hline
 rapidly create new colvars & colvar templates \\
 choose among available components  & link to list of components in online documentation \\
 rapidly define chosen components & component templates and online documentation \\
 define atoms involved in variables & insert atoms from representations, labeled atoms, or VMD selections \\
 double-check atom selections & show colvar atoms in molecular representation\\
 \hline
 explore dependence on atomic coordinates & colvar gradient visualization \\
 explore dependence on specific molecular motion & interactively move atoms/molecules and update colvars \\
 explore time-behavior of variables & timeline plots\\
 explore correlation between variables & pairwise scatterplots\\
 visualize specific high-dimensional quantites & rotation display, volume map display\\
 other quantitative analysis & export timeline to external software
\end{tabular}
\caption{Summary of useful tasks in a collective variable workflow, and the corresponding technical solutions implemented by the Colvars Dashboard}
\label{tab:tasks}
\end{table}

Internally, the Colvars Dashboard is written in the scripting language Tcl, and calls four programming interfaces:
\begin{enumerate}
 \item the Collective Variables Module, to create and query colvars;
 \item VMD, to drive the display of specific trajectory frames, interact with atom selections and representations, and visualize Cartesian gradients;
 \item Tk, to create the graphical user interface;
 \item the VMD plugin Multiplot, for interactive plots.
\end{enumerate}

The Collective Variables Module (\textit{Colvars}) is already included in official binary releases of VMD, and thus available to all users on Linux and MacOS platforms since version 1.9.2, and on Windows platforms starting with VMD 1.9.4.
The Dashboard graphical plugin is also distributed with VMD 1.9.4 and later (the recommended version as of publication is 1.9.4a54).
The latest version can be downloaded at any time from the Colvars public repository on GitHub (\url{https://github.com/Colvars/colvars/tree/master/vmd/cv_dashboard}), and updated in an existing VMD installation without recompiling.

\subsection{Underlying library: the Collective Variables Module, distributed with VMD}

\textit{Colvars}~\cite{Fiorin2013} is a portable C++ library, with an interface class \texttt{colvarproxy} that allows for writing new interfaces to C++, C, or Fortran programs without modifying the core of Colvars.
In MD simulation programs, Colvars is called at every timestep to update variables and biases as needed.
In contrast, in VMD, access to Colvars features is enabled \emph{on-demand}, through its rich Tcl scripting interface.
This interface, already available since the Colvars versions included in the VMD 1.9.3 release, allows for dynamically allocating the library, creating and deleting variables and biases, parsing configuration files, and computing quantities such as the variables' values and atomic gradients on the fly, using the current atomic coordinates stored by VMD in memory.
The Dashboard here introduced leverages this existing scripting interface, and binds the most commonly used functionality to simple graphical objects such as menus and buttons.

\subsection{Discoverability and documentation}

The Dashboard's graphical interface is designed to be simple and discoverable.
Rather than menus, all functions are visible as clearly labeled buttons, making the learning curve quite smooth.
Power users may use an extensive collection of keyboard shortcuts, some of which are directly mentioned in the interface, while all are included in the documentation.

The documentation is available online as part of the VMD edition of the Colvars user's guide.\cite{ug_dashboard}

The Configuration Editor window, the use of which requires the most background information, features direct links to the relevant sections of the online documentation.
The template configurations provides syntax examples for virtually all features, while the online documentation provides a reference and background explanations.

\subsection{Interaction with MD engines}

To ensure a smooth  workflow, the same Colvars configuration file should be used in sequence in the Dashboard and in MD simulations, without modifications.
The compatible MD engines are all those with functional interfaces with Colvars: LAMMPS \cite{Thompson2021}, NAMD \cite{Phillips2020}, and our public, patched version of GROMACS \cite{Abraham2015}.
Most of these programs use different unit systems from each other: in particular, LAMMPS is used for a broad variety of application domains and offers a choice of multiple unit systems that can be set by the user.
To enable interoperability with all MD engines, the Colvars/VMD interface performs unit conversion between VMD's internal units and the current set of units for Colvars.
The unit system can be changed within the graphical interface.
A ``units'' flag is added to the Colvars configuration to avoid accidental use of inconsistent units: running Colvars with a specific MD engine will trigger an error condition if the configuration specifies units that are inconsistent with those of the engine.

\section{Task 1: Visualizing a given set of variables}

\begin{figure}[!ht]
\centering
 \includegraphics[width=7cm]{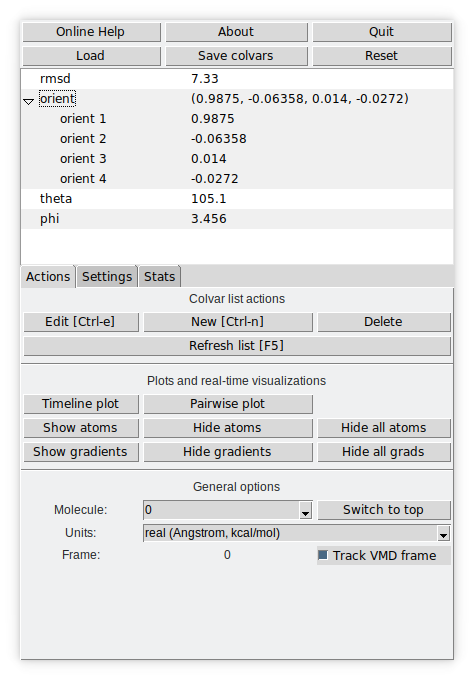}
 \caption{The Collective Variables Dashboard main window.}
\label{fig:main_window}
\end{figure}

The main window~\ref{fig:main_window} features a two-column table listing the variables on the left, and their values for the current timestep on the right.
Vector variables such as vector distances or orientation quaternions can be expanded to select their scalar components individually for plotting. 
By default, values are updated in real-time to track the current trajectory frame displayed by VMD, however this can be disabled especially for computationally expensive sets of variables that are not suitable for real-time display.

\subsection{Loading and saving configuration files}

Key to integration of the interactive colvar modeling process into the simulation workflow is the ability to read and write the same configuration files that would be used in biased simulations in any of the supported MD engines.
The selection of colvars that has been refined interactively can be saved at any time.
The Dashboard only handles collective variables themselves, not biasing methods.
therefore, the configuration of biases should be kept in a separate file and loaded separately in MD simulations engines.

\subsection{Visualizing trajectories simultaneously in VMD's molecular representations and in colvar space}

\begin{figure}[!ht]
\centering
 \includegraphics[width=7cm]{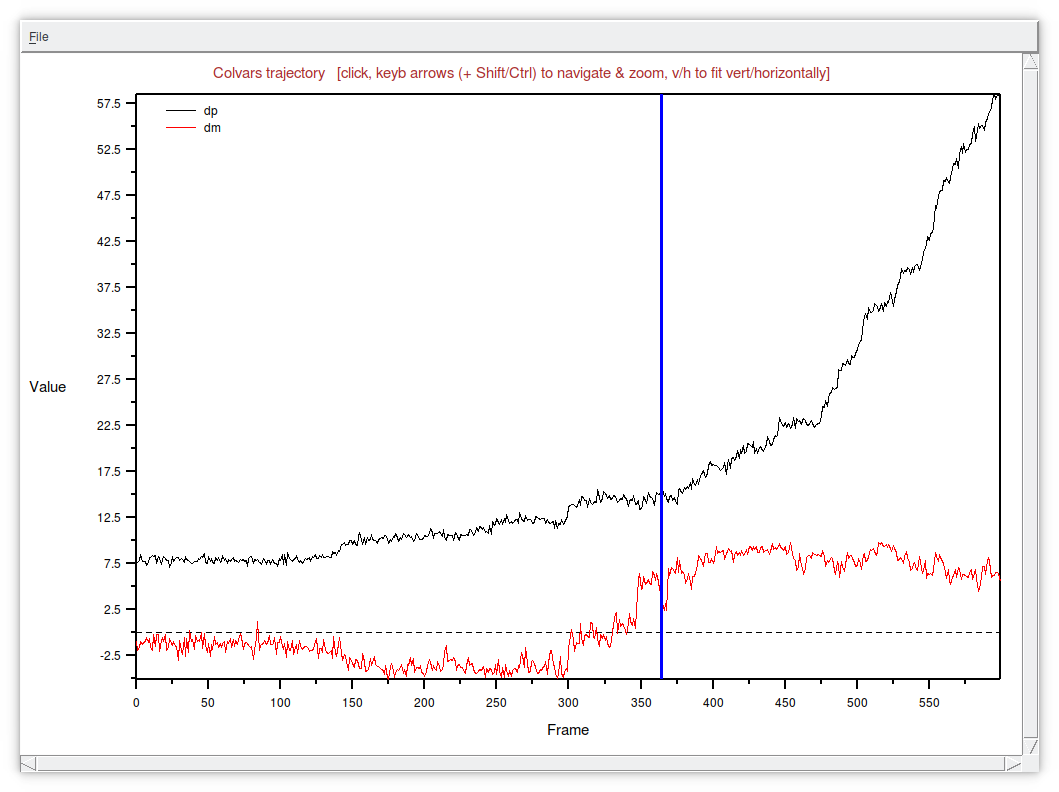}
 \includegraphics[width=7cm]{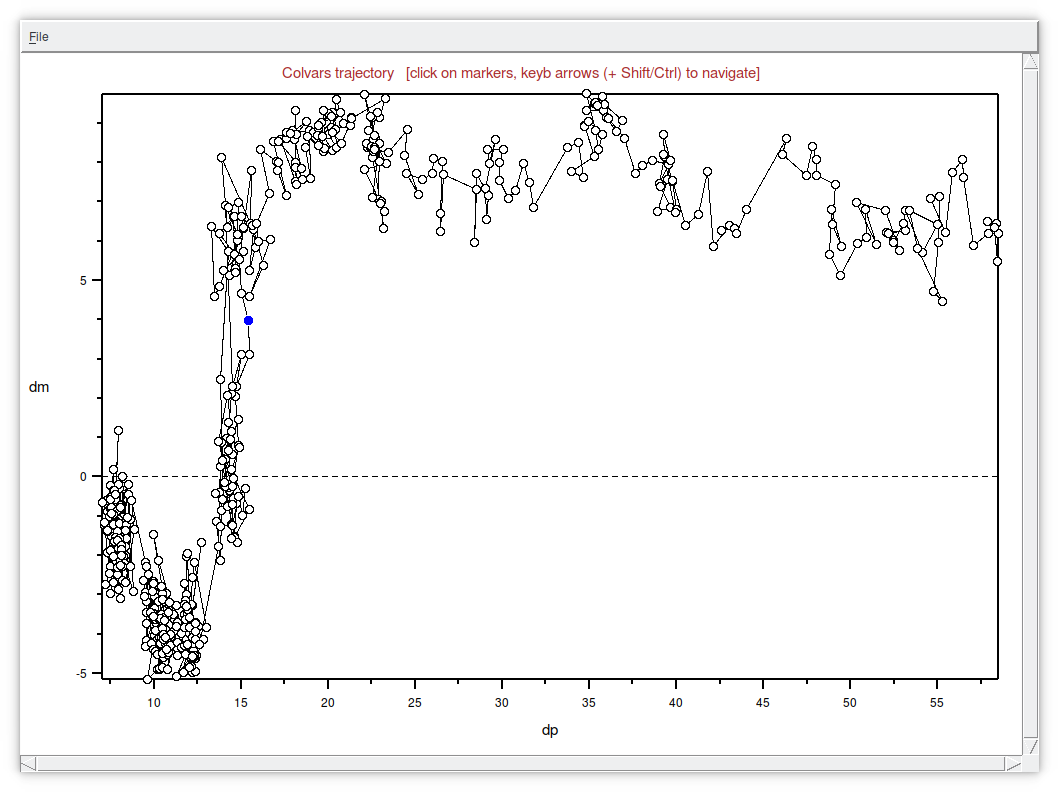}
 \caption{Plots of collective variable values along a loaded trajectory. Left: Timeline plot. Right: Pairwise plot. The vertical blue line in the timeline and blue dot in the pairwise plot indicate the current frame in VMD.}
\label{fig:plots}
\end{figure}

The timeline view provides a plot of the values of a selection of variables as a function of time (Figure~\ref{fig:plots}).
A vertical time cursor points to the current frame of the trajectory being displayed, and moves when the trajectory is animated.
Conversely, clicking on any point in the timeline window brings VMD's display to that frame in the trajectory.
Left and right arrow keys on the keyboard navigate in the trajectory, while up and down keys provide a zoom in/zoom out functionality to access detailed features of long trajectories.

The pairwise scatter plot can be created for pairs of scalar colvars (or scalar components of vector variables), and is useful for exploring correlation between variables (see Figure~\ref{fig:hsp_proj}).
As in the timeline view, trajectory animation is visualized, with a colored dot representing the current frame. A click on a dot brings the VMD representation to the frame in question.

\subsection{Exploring the dependence of colvars on atomic coordinates}

Atom coordinates can be modified interactively using VMD tools (translate and rotate atoms, residues, and molecules using a mouse or another device). Then the table of colvars can be updated almost instantly by pressing F5, allowing for a near-real-time exploration of changes in collective variables in response to changes in structure.

\begin{figure}[!ht]
\centering
 \includegraphics[width=5cm]{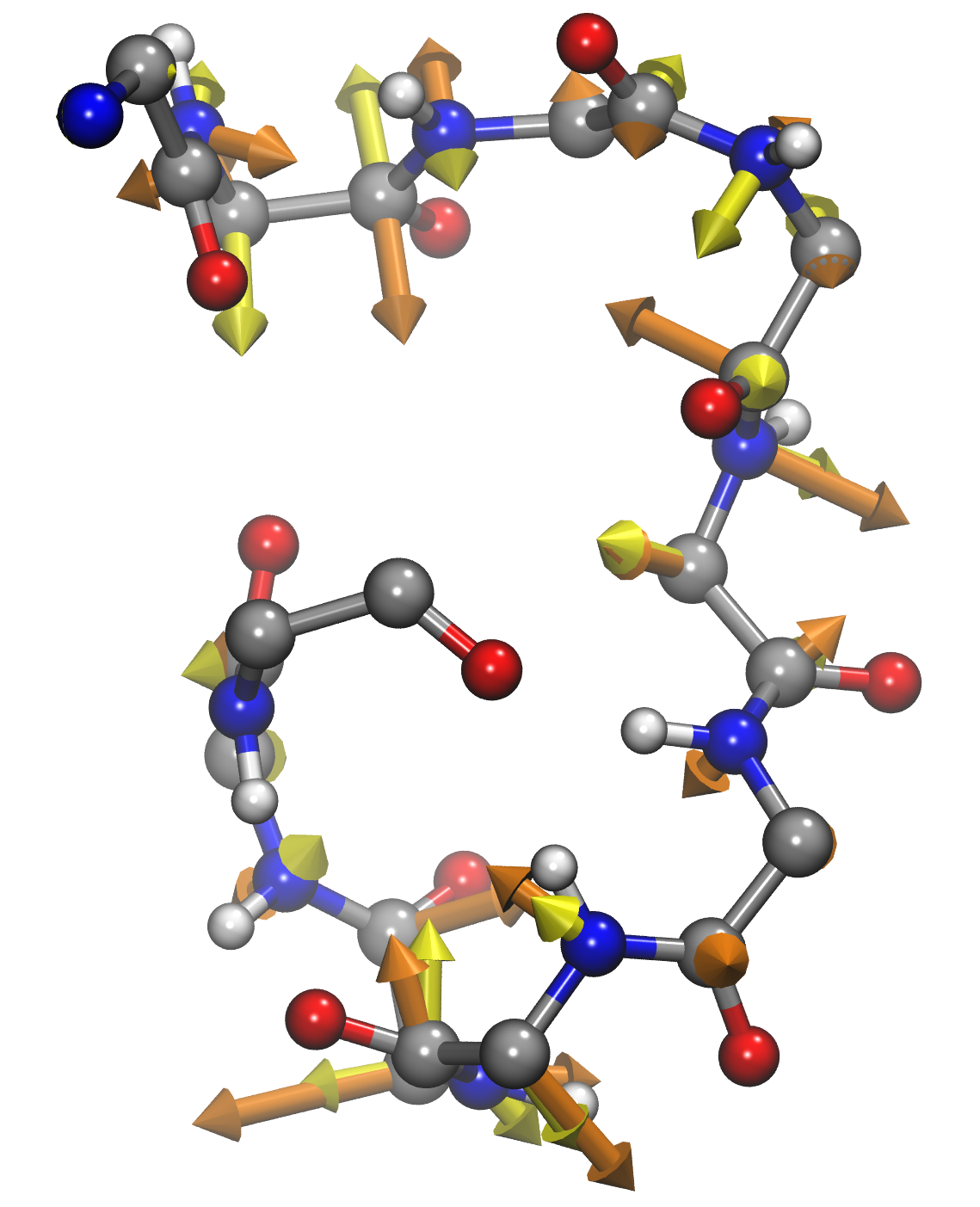}
 \caption{VMD molecular representation of the backbone of deca-alanine, featuring the atomic gradients of its first two dihedral principal components,\cite{Altis2007} as defined in reference~\cite{Henin2021}.
 Gradients of the the first and second mode are represented by yellow and orange arrows, respectively. The arrow length scale is arbitrary but identical for both coordinates, allowing for comparing the gradient magnitudes.
 }
\label{fig:gradients}
\end{figure}

In addition, the Dashboard features a real-time visualization of the gradient of an arbitrary scalar variable with respect to Cartesian coordinates (Figure~\ref{fig:gradients}).
In this visualization, the gradient of the variable with respect to the position of each atom is represented by an arrow starting from the atom in question.
Like the colvar values, the gradients are computed on the fly, and can be updated near-instantly after a change in coordinates.
This is particularly useful for variables whose mathematical expression is too complex to build a clear intuition, such as dihedral principal components~\cite{Altis2007} (Figure~\ref{fig:gradients}), or distance to bound configuration~\cite{Salari2018} (Figure~\ref{fig:DBC_grads}).

\subsection{Dedicated visualizations for specific quantities}

\begin{figure}[!ht]
\centering
 \includegraphics[height=4cm]{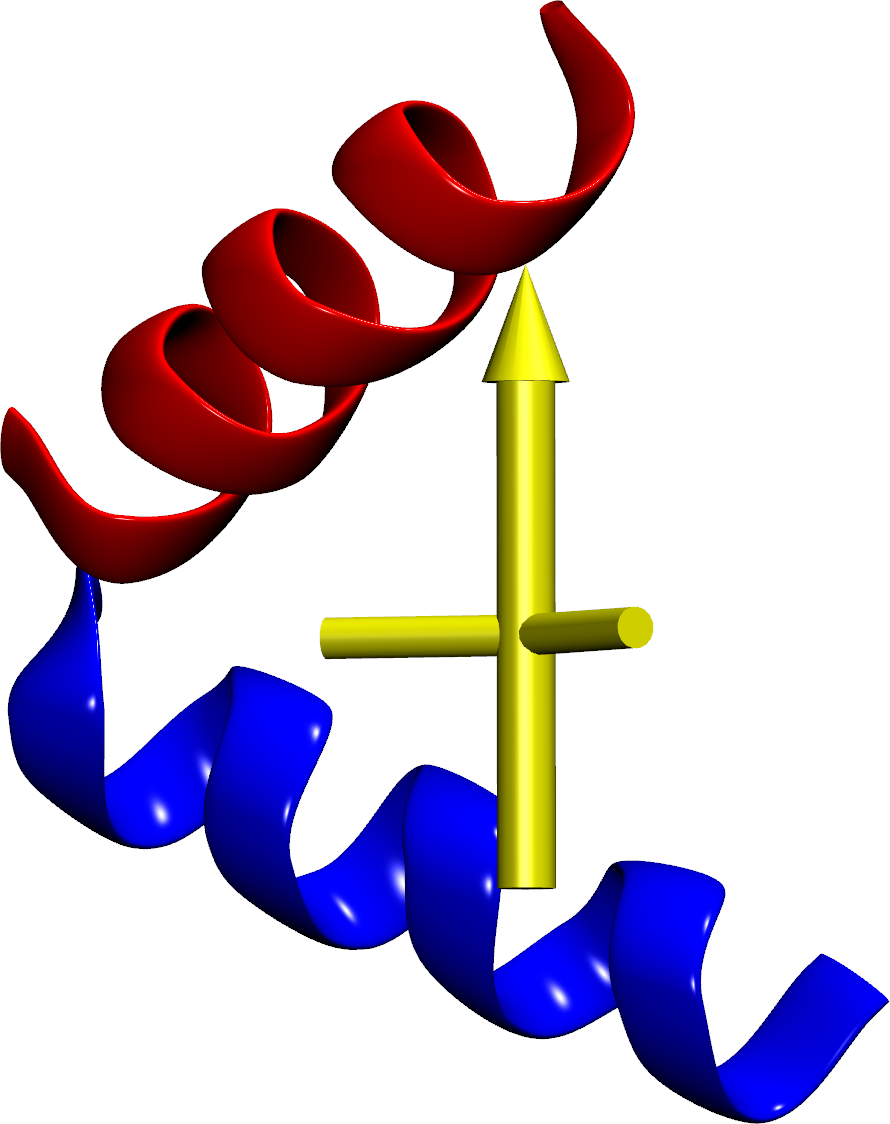}
 \includegraphics[height=4cm]{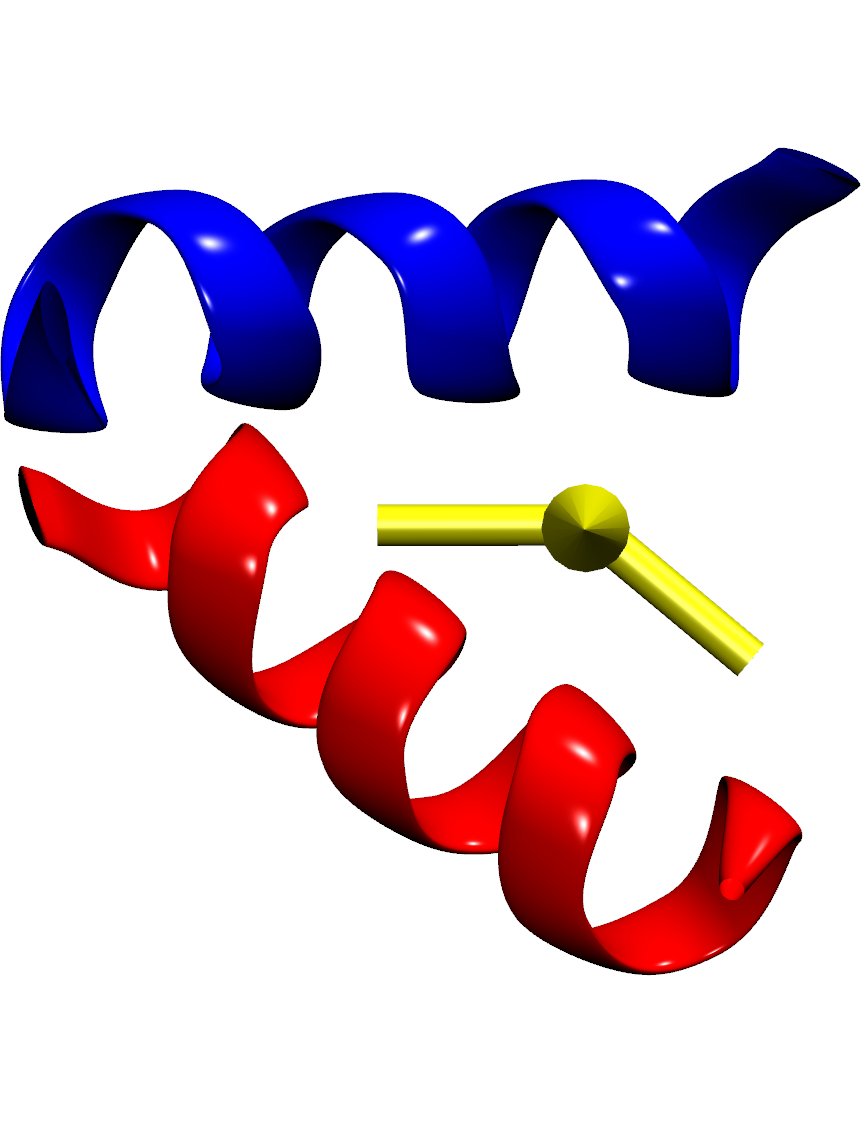}

 \caption{Graphical representation of a rotation operator characterizing the orientation of a molecule with respect to a reference, viewed orthogonal to (left) and along (right) the axis of rotation.
 Here the molecule of interest (red) and the reference (blue) are two copies of a 15-alanine helical peptide. The rotation operator is represented by the yellow arrow, oriented along the axis of direct rotation, and the angle between the two side ``arms'' is the angle of rotation.}
\label{fig:rotation}
\end{figure}

Colvars offers a special type of variable to represent rotation operators, encoded as unit quaternions.
In particular, the \texttt{orientation} component describes the collective rotation of a group of atoms with respect to reference coordinates, computed based on a roto-translational least squares fit\cite{Fiorin2013}.
This is particularly useful to capture the collective rotation of a flexible object in a robust way.
Rotations have three degrees of freedom (e.g.\ two defining the axis and one defining the angle of rotation), and are difficult to visualize in an intuitive way.
The Dashboard offers a specific representation for such variables, illustrated in Figure~\ref{fig:rotation}.
An arrow represents the axis of rotation, oriented so that the rotation angle is between 0 and 180 degrees in the direct (counterclockwise) sense.
The angle itself is represented by two segments protruding from the axis.

The Dashboard also facilitates the visualization of volume maps associated with existing colvars: see Section~\ref{sec:multimap} for details.

\section{Task 2: Designing new collective variables}

\subsection{Editing colvar configurations}

The Dashboard includes a Collective Variable Editor dialog, which presents the text of the configuration string, as well as a number of helper features that make creating and modifying variables much faster and less error-prone.
The editor contains direct links to relevant sections of the online documentation.
To soften the learning curve, most of the features can be inserted as templates which require only limited modifications.
In addition, possible syntax errors are handled gracefully:
while the configuration of an existing variable (or several variables) is being edited, its previous working configuration is backed up.
If the user submits an edited configuration string and Colvars returns an error, the complete error message is displayed, and the editor window is shown again, providing a chance to fix the error.
If the window is closed before any changes can be applied successfully, the saved, functioning version of the variable(s) is restored.

\subsection{Inserting atom selections}

A significant part of defining collective variables is to specify which atoms are involved in the various groups used to calculate the variable.
The Dashboard provides several helper features to create atom selections:
\begin{enumerate}
    \item From a VMD atom selection text. This helps leverage the very powerful atom selection language of VMD. Furthermore, atom selections created this way are dynamic: the selection text is embedded in the Colvars configuration string as a special comment, and recomputed every time the configuration is parsed, possibly adapting to changes in coordinates, or topology if the same configuration is applied to different molecules.
    \item From a VMD representation, selected via a dropdown menu. This lets the user immediately transpose atom groups being visualized into collective variables.
    \item From "picked atoms" selected using the mouse in VMD's graphical window.
\end{enumerate}

The last helper is a file picker dialog, which allows for inserting the paths of files containing reference coordinates or atom group definitions, together with the relevant keywords.

\begin{figure}[!ht]
\centering
 \includegraphics[width=10cm]{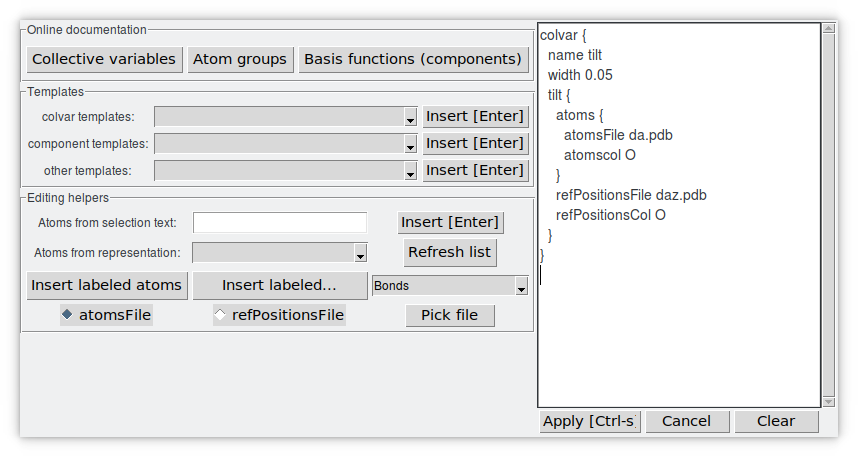}
 \caption{Colvar configuration editor window.}
\label{fig:editor}
\end{figure}

Atom groups used by existing variables can be visualized by selecting them and clicking ``Show Atoms'' in the main window.
This creates VMD representations showing atoms as spacefill, with distinct colors for each atom group.
This is useful to check that the designated atoms correspond to those intended, preventing issues such as off-by-one errors due to conflicting atom numbering schemes between software.
Note this is the reverse operation to creating an atom group directly from a representation.

\subsection{Inserting templates}

The editor interface allows for inserting templates for all types of basis functions, and for colvars with specific options whose syntax might be difficult to recall.
The templates contain placeholders for key parameters, and comments that function as a first level of inline documentation.
More in-depth documentation is available by clicking the documentation links.

\subsection{Inserting geometric observables from VMD}

VMD lets users select and measure distances (\textit{Bonds}), angles, and dihedral angles between particles by selecting them with the mouse in the graphical window.
These geometric quantities can be directly inserted as collective variable component definitions.

\subsection{Setting up biased MD simulations}

Typically, a project will start with unbiased simulations for an agnostic, local exploration of configuration space.
The Dashboard greatly reduces the time needed to test new ideas for collective variables based on any such trajectories, biased or unbiased, to use in further biased simulations.
Although major MD engines support ``reruns'', analysis of colvars in VMD is much quicker, and the new variables are evaluated frame by frame in real time.
This is even more relevant when analyzing simulations made with one MD engine but planning simulations with another one to take advantage of unique features in each package.
The Colvars Module allows this transferability, and the Dashboard facilitates it.
In particular, changes in unit systems between different MD engines are handled gracefully and safely.

\section{Example applications}

\subsection{Exploring ligand binding and unbinding}

\subsubsection{Monitoring ligand unbinding simulations}

\begin{figure}[!ht]
\centering
 \includegraphics[width=0.47\linewidth]{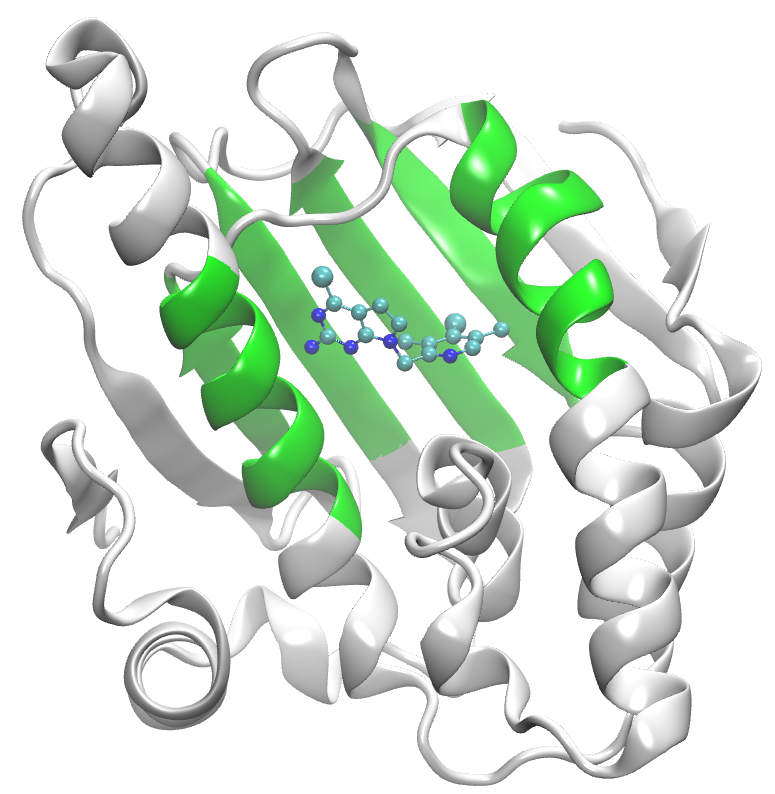}
 \includegraphics[width=0.42\linewidth]{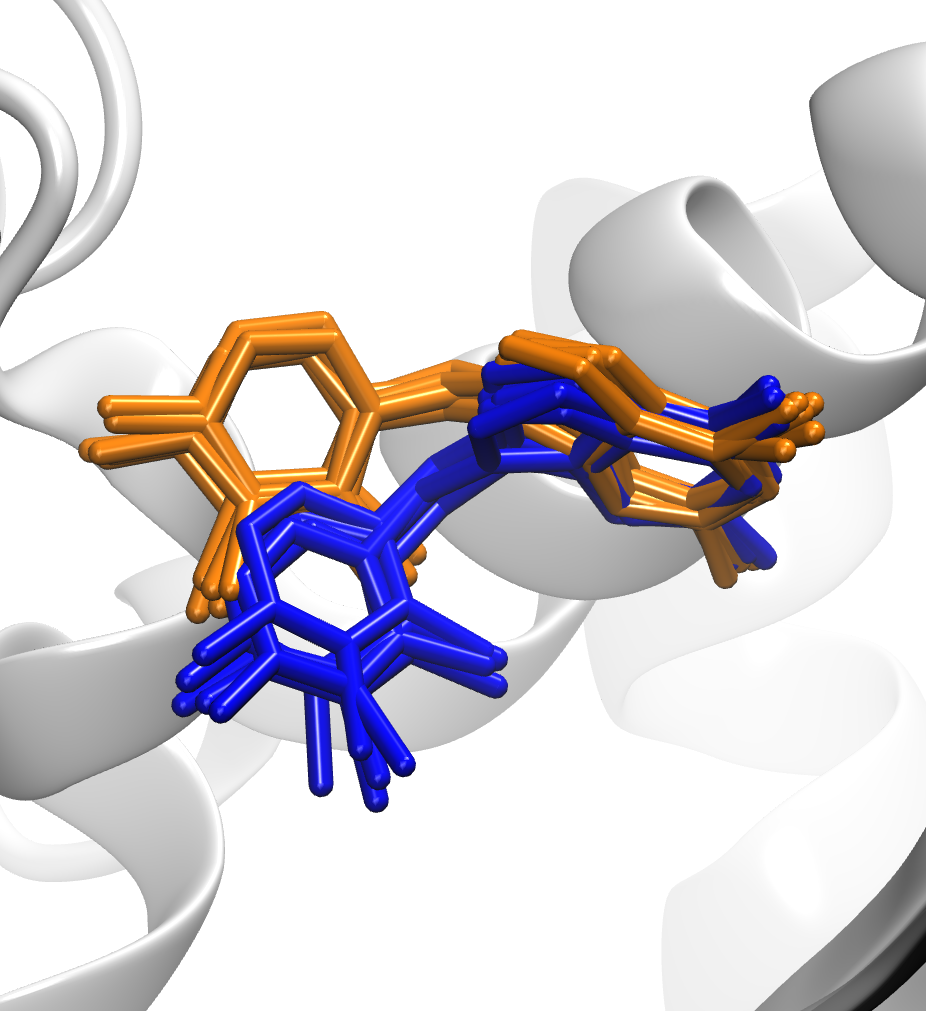}
 \caption{Left) The Hsp90/ligand system. The ligand is represented as ball-and-stick, and the protein as cartoon, with residues chosen to represent the binding site colored green. Right) In a zoomed view of the binding site, the two bound states are represented by a collection of configurations from simulations, shown as sticks: the main bound state is colored blue, and the secondary one is colored orange. The latter was discovered by analyzing ABMD trajectories.}
\label{fig:hsp_states}
\end{figure}

Quantitative studies of ligand unbinding kinetics raise major sampling challenges, which can be alleviated through the use of a number of enhanced sampling methods.
Many of those require a precise definition of the bound state as well as a suitable coordinate to track the ligand’s progress. In those tasks, the use of a tool like the Dashboard enables the user to test multiple colvar candidates in order to find what best suits the problem.
The example discussed in this section consists in the unbinding of a pharmacological ligand from the N-terminal domain of the heat shock protein 90 (Hsp90) chaperone (Figure \ref{fig:hsp_states}).
The system was first explored using Adiabatic Bias MD (ABMD)\cite{Marchi1999} (implemented in Colvars/NAMD), which generated a family of possible exit trajectories under a mild non-equilibrium bias.
Then an unbiased sample of exit trajectories as obtained using the Adaptive Multilevel Splitting (AMS) method\cite{Cerou2019}.
The Colvars Dashboard was used in the analysis of trajectories generated with both methods.

\begin{figure}[!ht]
\centering
 \includegraphics[width=0.49\linewidth]{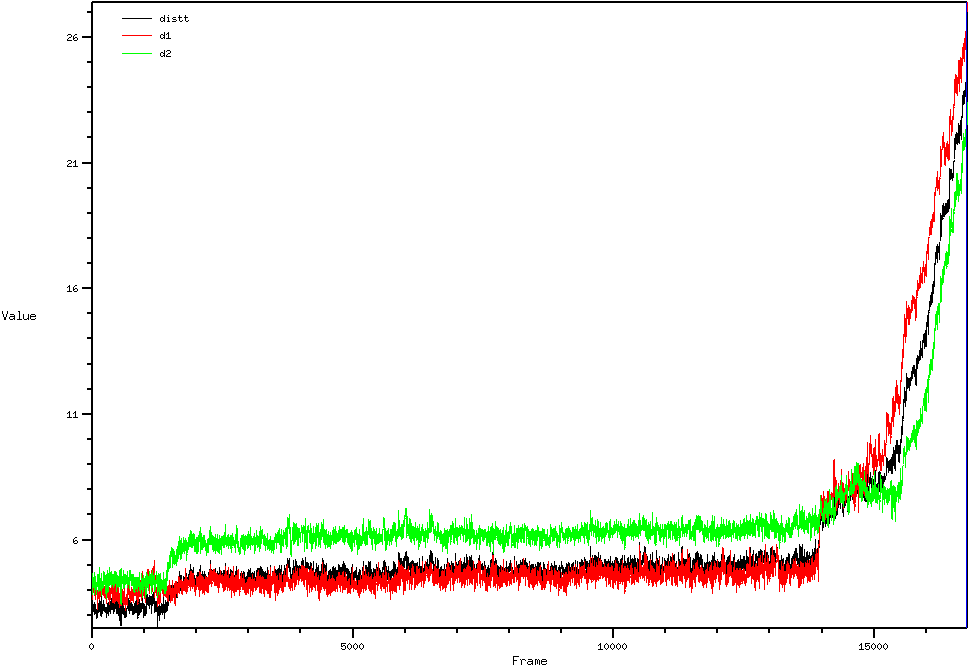}
 \includegraphics[width=0.49\linewidth]{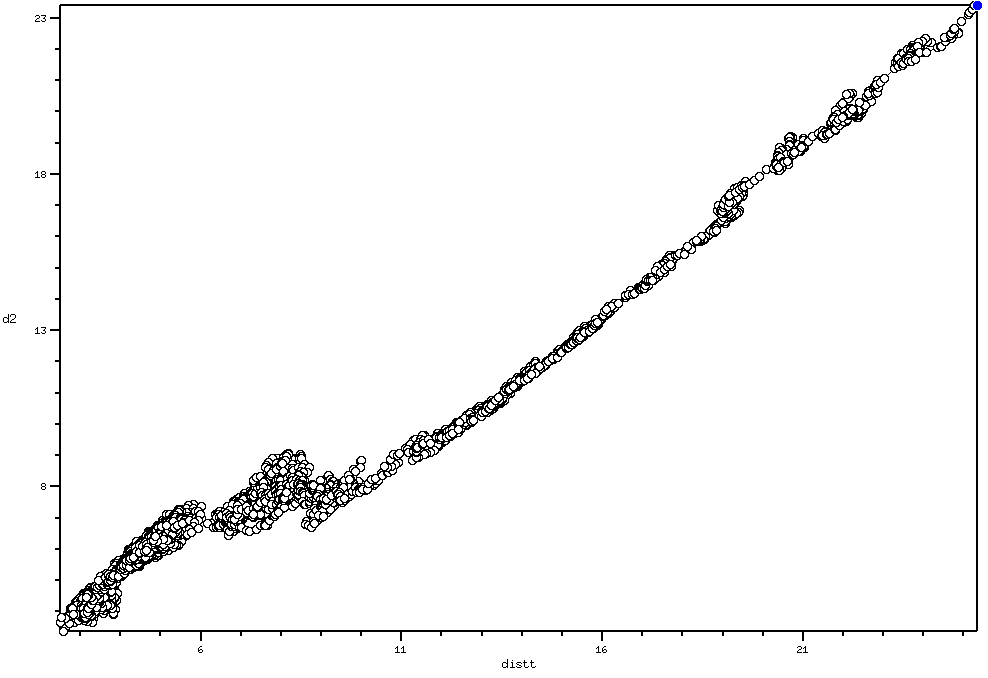}
 \caption{
 Colvars visualization of ABMD escape trajectories using the Dashboard. Left: Timeline of the center of mass distance to the protein cavity (black), the distance of the bicycle (red) and the distance of the aromatic ring (green).
 Right: Pairwise plot that correlates the global site-ligand distance \texttt{distt} with the distance from the aromatic ring \texttt{d2}}
 \label{fig:hsp_dists}
\end{figure}

The analysis of the ABMD trajectories using VMD reveal that the ligand detaches from the protein first by its aromatic ring, followed by its bicycle structure (Figure \ref{fig:hsp_states}). This suggests the distances from those two sections of the ligand should be considered separately. Figure \ref{fig:hsp_dists} shows the Dashboard session used in the exploration of new collective variables based on that idea. The time plot confirms that the aromatic ring is the first to disconnect from the protein, but also shows that this represents an intermediate state, as the ligand stays a while bound only through its bicycle. 

\begin{figure}[!ht]
\centering
 \includegraphics[width=0.28\linewidth]{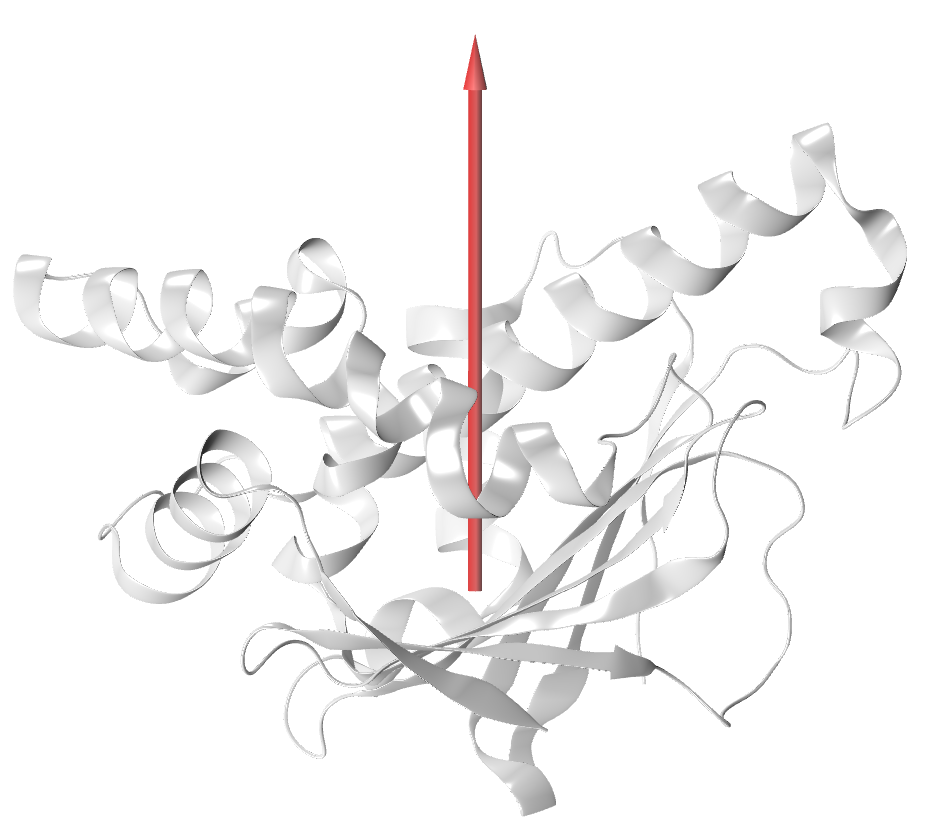}
 \includegraphics[width=0.35\linewidth]{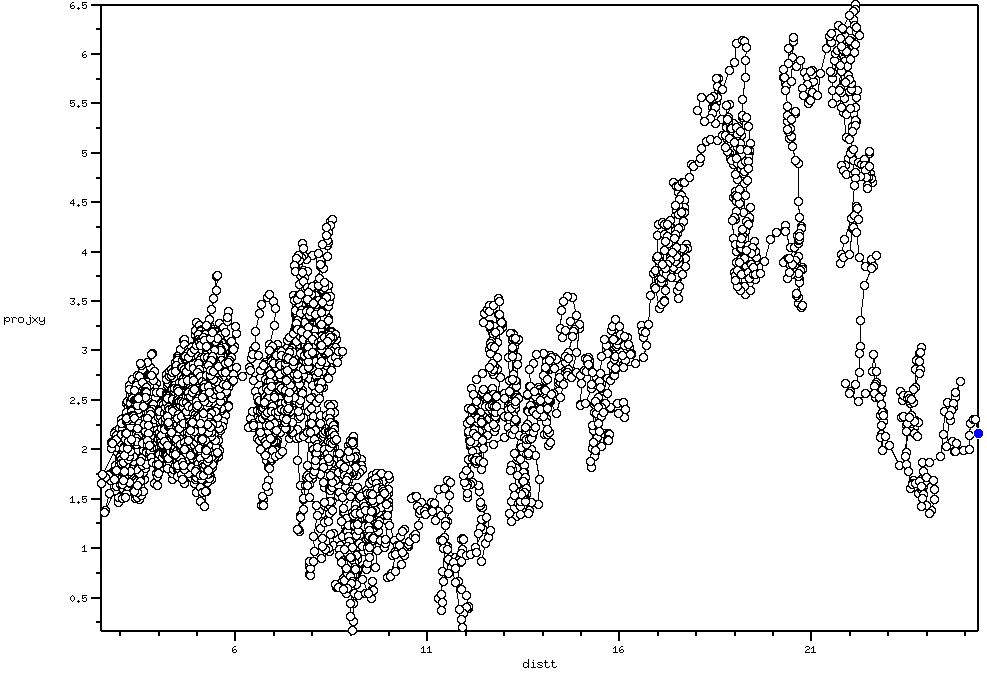}
 \includegraphics[width=0.35\linewidth]{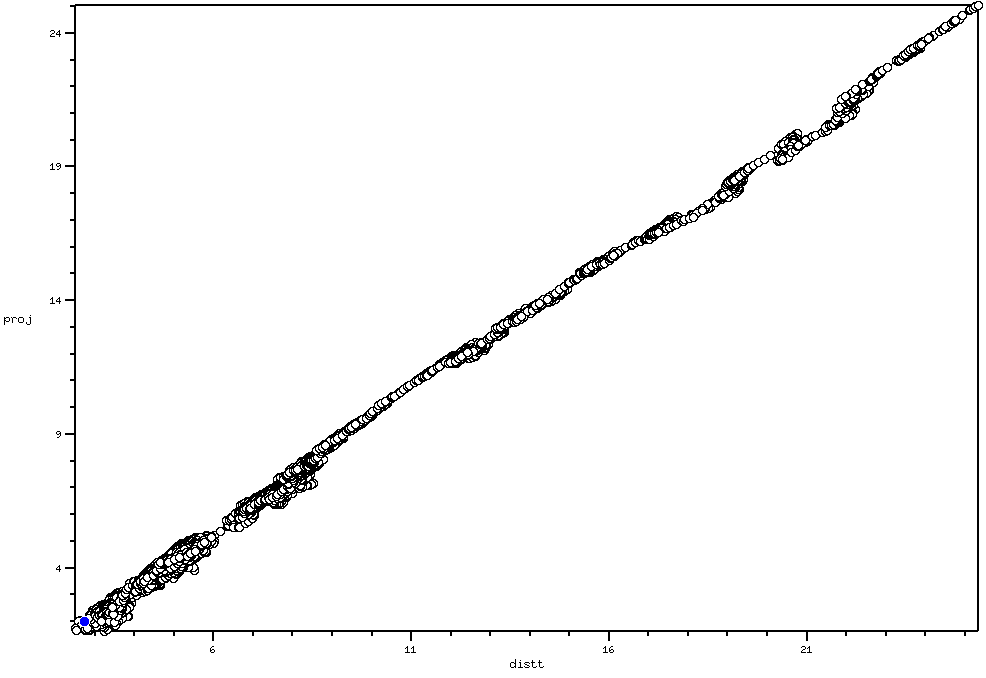}
 \caption{The axis (in red) used in the new explored coordinate. Both pairwise plots show the correlation between the distance to the cavity and its projection onto, respectively, the plane parallel to the cavity opening (left), and the axis perpendicular to it (right). The latter exhibits a high correlation with the isotropic distance in this trajectory.}
 \label{fig:hsp_proj}
\end{figure}

Another result from the analysis of those trajectories using Colvars Dashboard is that only one component of the distance vector is necessary to describe the ligand's exit. Figure \ref{fig:hsp_proj} shows the axis used in this new coordinate, along the direction of the cavity opening.
The pairwise plot between the distance and the projection onto the plane shows a poor correlation, but the pairwise plot with the projection onto the axis exhibits a high correlation, suggesting this new variable as a suitable progress coordinate.
Furthermore, compared with the isotropic distance, it provides a better resolution between states where the ligand is at a small distance from the protein and escaping along the axis leading away from the site, and those where the ligand is interacting nonspecifically with the protein surface outside the cavity.


The two trajectories generated with the AMS method exhibit a long residence time within two distance intervals distinct from the initial bound state, suggesting the existence of two additional metastable states. Because those intervals did not coincide but did overlap, it was necessary to find a new coordinate that could separate them, to distinguish them within the AMS approach.

\begin{figure}[!ht]
\centering
\begin{minipage}{0.39\linewidth}
 \includegraphics[width=\linewidth]{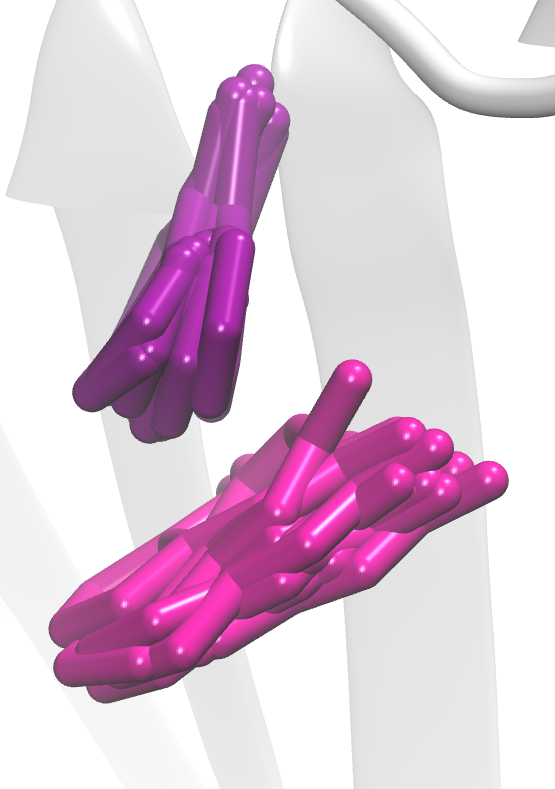}
 \includegraphics[width=\linewidth]{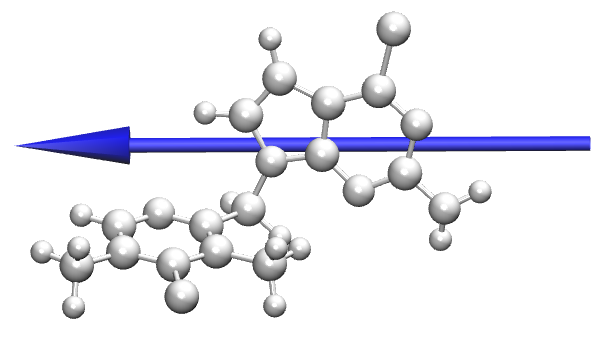}
\end{minipage}
\begin{minipage}{0.59\linewidth}
 \includegraphics[width=\linewidth]{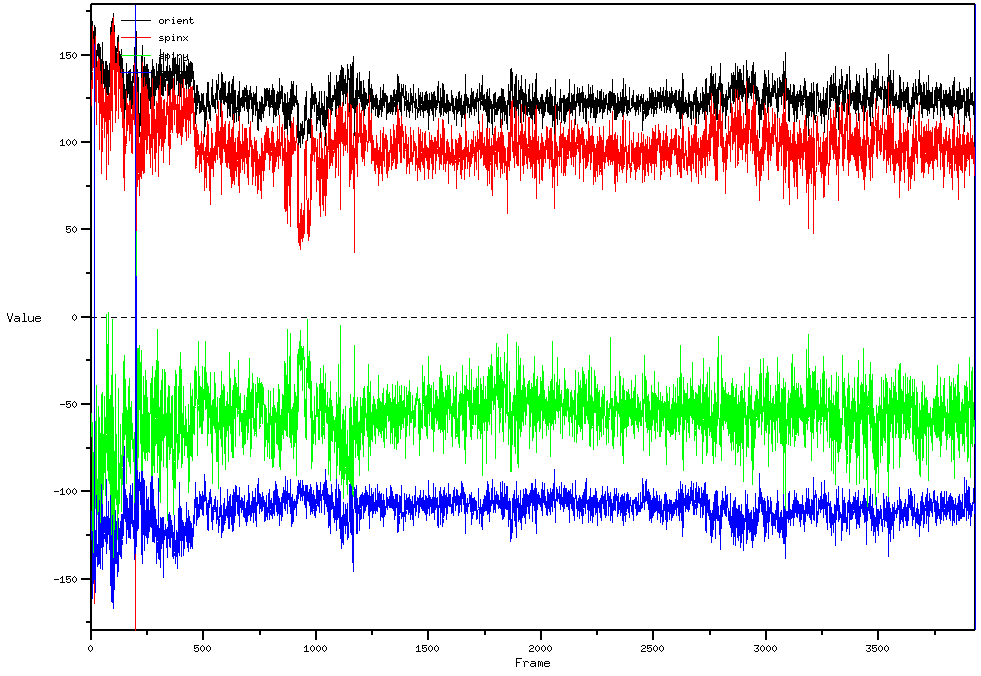}
 \includegraphics[width=\linewidth]{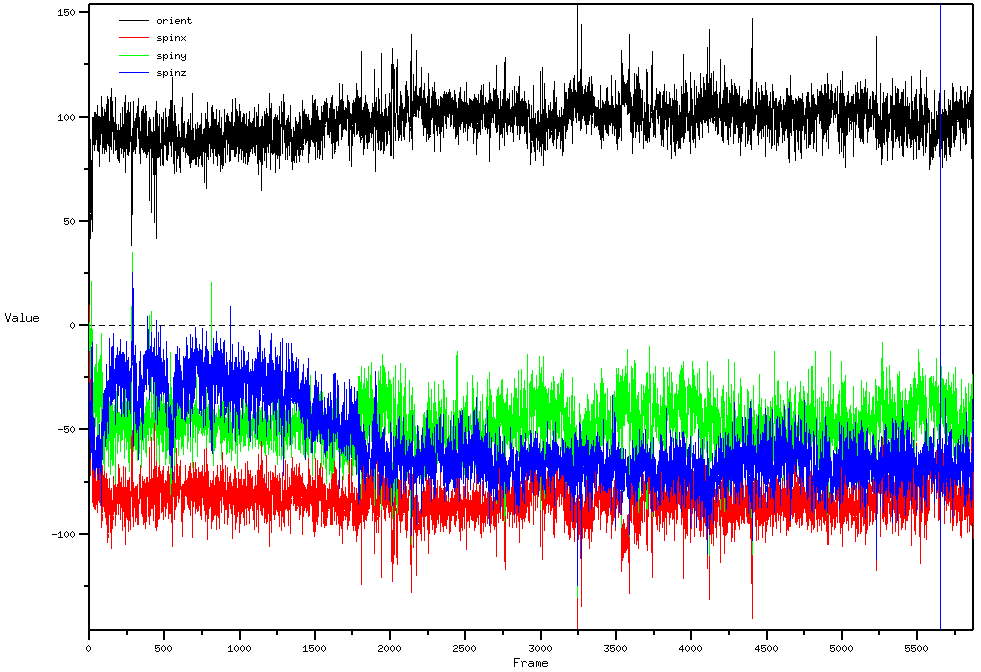}
\end{minipage}
\centering
 \caption{Left) Positions of the ligand's bicyclic moiety for the two AMS trajectories. The two modes differ by a rotation. At the bottom the axis used for the coordinate candidate (\textit{spinx}). Right) Timeline plots of 4 candidate coordinates along trajectories sampling each of the two states (upper and lower panel, respectively). The angle of rotation around the x axis (\textit{spinx}, red curve) proved a solid coordinate to separate the two states.
 }
 \label{fig:bicycle_angle}
\end{figure}

A rapid visual inspection of the trajectories suggested that the larger, bicyclic moiety of the ligand experiences a rotation between the two modes (Figure~\ref{fig:bicycle_angle}).
The Dashboard allowed for quickly testing rotation around different axes to separate those modes. Figure \ref{fig:bicycle_angle} show plots for the global rotation angle from the reference (\texttt{orientationAngle} component), and angles of rotation around selected axes (\texttt{spinAngle} component) \cite{Fiorin2013}.
The time plot of the different trajectories shows that the rotation angle around the x axis (in blue in Figure \ref{fig:hsp_proj}) was successful in separating the states.
These orientations of the ligand were computed relative to the protein, using the moving frame of reference formalism inherent to Colvars.\cite{Fiorin2013}

In this example the use of Colvars Dashboard was key in the exploration of unknown metastable states, the design of a better suitable progress coordinate, and the separation of metastable states. In the latter, the comparison between different coordinates was critical, as the relevant axis of rotation could not be determined by eye. Finally, the Colvars Dashboard was able to aid in the setting of new unbinding simulations, both in the progress coordinate and in the coordinates used to define different bound states.

\subsubsection{DBC coordinate for ligand binding}

Computational prediction of absolute binding affinities can also require careful design to ensure convergence, including a potentially complex ensemble of binding restraints connected to the definition of the binding site.\cite{duboue2021building}
The Streamlined Alchemical Free Energy Perturbation (SAFEP) approach\cite{Salari2018} offers a simpler protocol based on flat-bottom restraints on a single, scalar binding coordinate named  distance-to-bound-configuration (DBC).
DBC is a ligand RMSD with respect to an ideal bound pose, computed in a moving frame of reference tied to the binding site.\cite{Salari2018, Ebrahimi2021}
DBC captures the ligands conformational changes as well as its translational and rotational motion with respect to the binding site.
Crucially, DBC is orthogonal to all internal degrees of freedom or the receptor, so that restraint forces on DBC do not bias the receptor's internal dynamics.

\begin{figure}[!ht]
\centering
 \includegraphics[width=10cm]{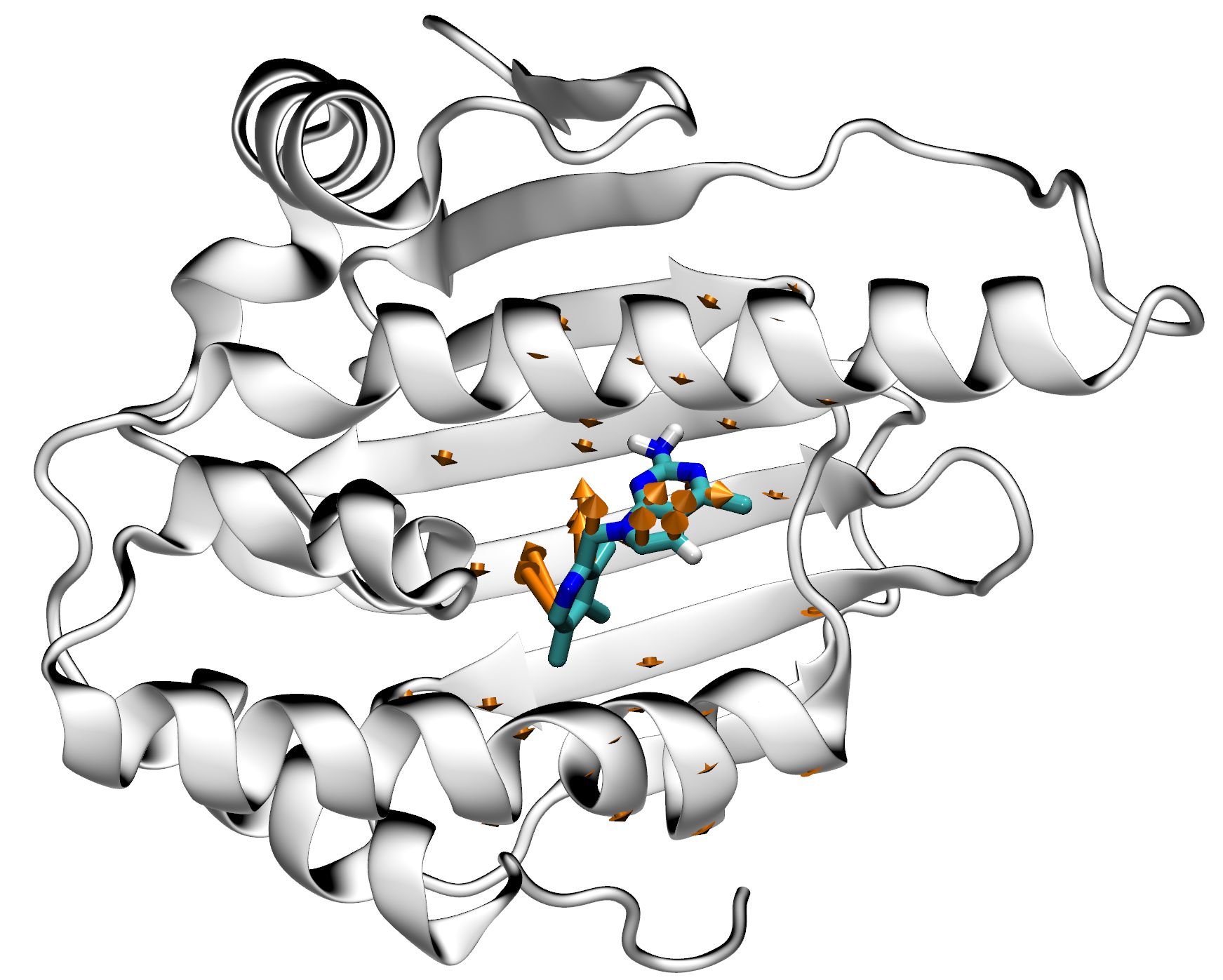}
 \caption{Gradients of the Distance-to-Bound-Configuration (DBC) coordinate visualized on a bound pose of Hsp90 and a pharmacological ligand. The arrows represent gradient components on atoms of the ligand and those protein atoms involved in the binding site definition. The gradient components on ligand atoms represent roto-translational as well as conformational degrees of freedom, whereas the components on protein atoms correspond to the purely roto-translational, rigid-body motion obtained by least-squares fit.\cite{Fiorin2013, Salari2018}}
\label{fig:DBC_grads}
\end{figure}

As the DBC coordinate aggregates many degrees of freedom, it is helpful to make its meaning more intuitive by visualizing its gradient for different ligand configurations.
Close to the reference bound pose, DBC is mostly sensitive to the ligand conformation and orientation (as in Figure~\ref{fig:DBC_grads}), while at larger values, it reflects mainly global translation away from the site.

Multimodal binding requires particular care to be correctly described, especially in quantitative affinity predictions.
\begin{equation}
    e^{-\beta \Delta G} = \sum_i  e^{-\beta \Delta G_i} \ ,
\end{equation}
where $\Delta G$ is the global effective binding free energy,  $\Delta G_i$ is the free energy of binding to individual mode $i$, and $\beta = 1 / k_B T$.
Different modes in the DBC distribution (or those of other descriptors) can easily be related to different binding modes using the timeline view of the Dashboard.

\input{multimap}

\section{Conclusion}

The Colvars Dashboard of VMD constitutes a new type of interface to the versatile Collective Variables Module, making it more useful for the purpose of fostering intuition, while easing its learning curve.
The interface is designed to be both accessible to newcomers, and very effective for expert users.
It facilitates the definition of new variables from the wide repertoire afforded by the Module, and their visualization, tightly integrated with the molecular graphics of VMD.
The Dashboard also seamlessly integrates with biased simulations using any of the molecular simulation software interfaced with the Collective Variables Module.
We expect that the Dashboard will help turn collective variable analysis from a somewhat specialized computational method to an accessible tool for understanding any simulated molecular trajectory.

\section{Acknowledgments}
JH acknowledges support from the French National Research Agency under grant LABEX DYNAMO (ANR-11-LABX-0011), and  from the Laboratoire International Associé CNRS/UIUC.

\bibliographystyle{unsrt}
\bibliography{colvars}

\end{document}

%% file: multimap.tex
\subsection{Volumetric maps and Multi-Map variables}
\label{sec:multimap}

\subsubsection{Support for volumetric maps by the Colvars module and the Dashboard}

When using enhanced-sampling MD in biophysical applications, two frequent requirements are the ability to tackle complex changes in macromolecular shape, and the ease of implementation of a suitable CV.
One approach aimed at attaining both is defining a CV as a function of a \emph{volumetric map}, expressed on a three-dimensional grid of Cartesian coordinates.
Additionally, the use of multiple maps to represent the distinct relevant states of one macromolecule (e.g.\ a protein) or many molecules (e.g. a lipid membrane) allows for reliable thermodynamic sampling, a capability provided by the Multi-Map method.\cite{Fiorin2020}

The Colvars Dashboard is uniquely useful in assisting with the preparation and the analysis of volumetric map and Multi-Map collective variables.
VMD already provides functionality to analyze and manipulate atomic structures alongside volumetric maps, a feature that has been exploited previously in the framework of the MDFF structural refinement protocol.\cite{Trabuco2008,Singharoy2016}
Recently, the Colvars module has been extended with the ability to access directly the volumetric maps loaded in VMD's memory, and compute CVs based on volumetric maps, such as ``Multi-Map'' variables.\cite{Fiorin2020}
The Colvars Dashboard leverages this capability to assist the two Tasks listed above, allowing to compute volumetric map-based CVs within VMD.
This can be particularly advantageous when the maps themselves are being prepared in VMD, such as the case of Multi-Map CVs based on protein density maps, as exemplified in ref.~\cite{Vant2020}.

Additionally, the Dashboard's graphical user interface allows for one-click selection and visualization of the maps themselves, consistent with how groups of atoms and gradient vectors are also selected and visualized (see Table~\ref{tab:tasks}).
Figure~\ref{fig:map-cv-eq} shows molecular renders including graphical representations of the volumetric maps generated through the Dashboard.
This is achieved in a convenient manner by querying VMD for the minimum and maximum values of each map, and using the mid-point between them to define a reasonable contour level (of course, such choice may be customized per user preference).

\subsubsection{Example application: wetting/dewetting of a region confined in a protein's structure}

Beyond inducing shape changes in membranes\cite{Fiorin2020} or proteins\cite{Vant2020}, a useful biological application of volumetric map-based CVs is enhancing the sampling of different wetting states in a confined protein cavity.
A simplified example of this was already shown in ref.~\cite{Fiorin2020}, using a single-wall carbon nanotube in water as a minimalistic system.
Here, to illustrate the use of map-based CVs on a biologically significant system, we consider the transmembrane pore of the M2 proton channel of the influenza virus.
The pore of this channel contains several clusters of water molecules, many of which resolved crystallographically\cite{Acharya2010}, in dynamic exchange with each other and with the outer water buffer.
Using a previously equilibrated snapshot of the protein embedded in a lipid bilayer\cite{Gianti2015} and following an additional 50~ns of equilibrium MD simulation at 300~K and 1~atm, the spatial distribution of water molecules in the pore region was examined a different levels of resolution.

\begin{figure}[h]
  \centering
  \includegraphics[width=\textwidth]{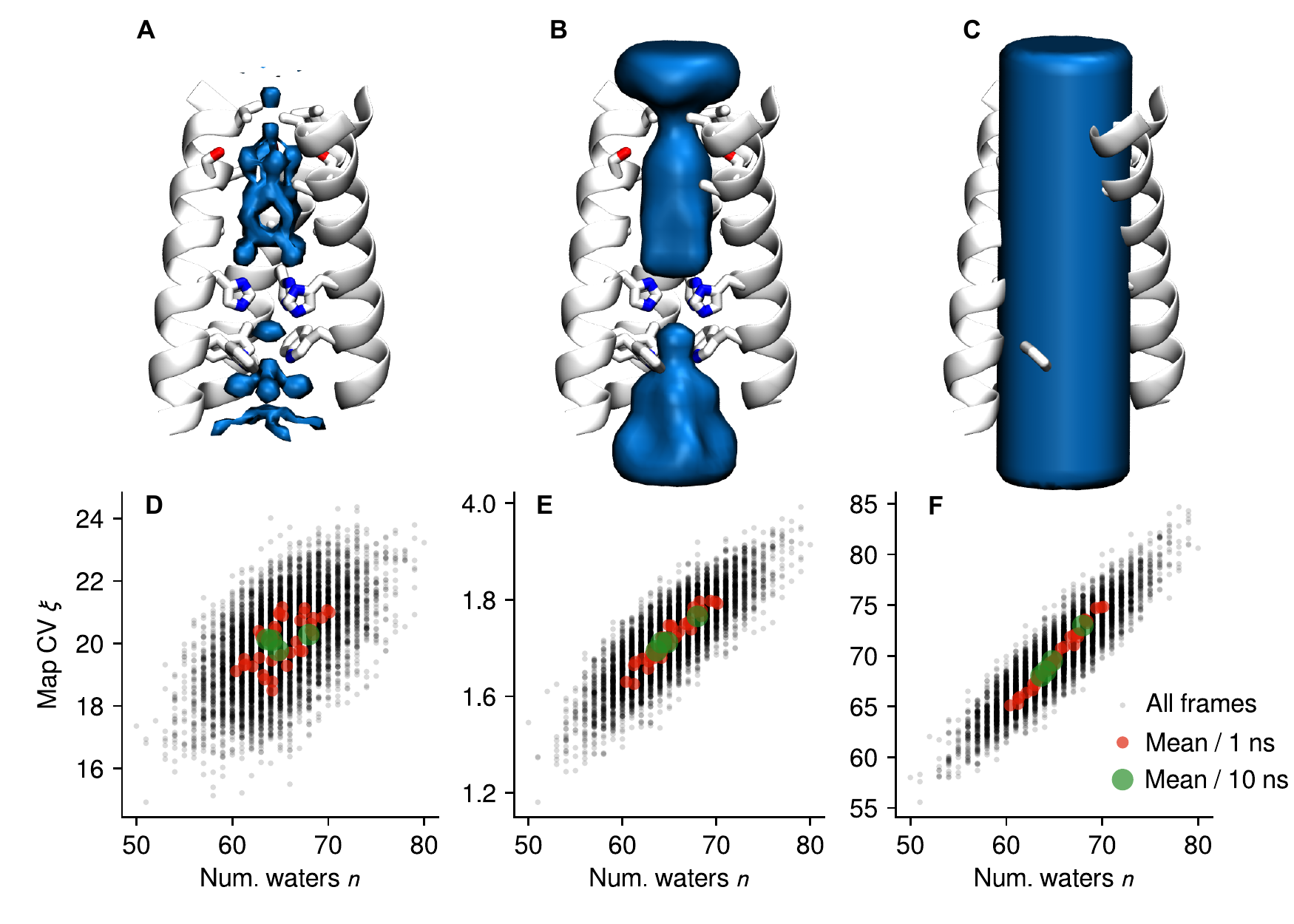}
  \caption{Shown are the M2 transmembrane channel (shown as ribbons) as well as three volumetric maps containing respectively: \textbf{(A)} the unsmoothed histogram of oxygen atoms from equilibrium MD simulation, \textbf{(B)} the smoothed histogram of the same, and \textbf{(C)} a smoothed step function defined on a cylindrical region; each contour is drawn at 50\% of the map's maximum value.
    Also shown in panels \textbf{(D-F)} are the values of the CVs based on the same volumetric maps, computed in VMD using the Colvars Dashboard and plotted as a function of the number of water molecules in the same selection.}
  \label{fig:map-cv-eq}
\end{figure}

Three different representations of the water density in the pore were used to generate map-based CVs for enhanced sampling.
First, the distribution of water oxygen atoms in a cylindrical region of 6~\AA{} radius and 40~\AA{} length was computed using VMD's \texttt{volmap occupancy} command (Fig.~\ref{fig:map-cv-eq}A).
Second, the same distribution was convoluted to a Gaussian kernel with half-width $\sigma$ $\simeq$ 1.5~\AA, using the \texttt{volmap density} command (Fig.~\ref{fig:map-cv-eq}B).
Third, a smoothed step-function was used to define a volumetric map equaling 1 within the cylindrical region described above and zero outside of it; a smooth transition was modeled as the right branch of a Gaussian function (Fig.~\ref{fig:map-cv-eq}C):
\begin{equation}
  \label{eq:smoothed_step}
  \phi(r) = \left\{
    \begin{array}{l l}
      1 & \mathrm{if }\ r \leq{} r_0 \\
      \exp\left(-\frac{(r-r_0)^2}{{\sigma_r}^2}\right) & \mathrm{if }\ r > r_0 \\
    \end{array}
  \right.
\end{equation}
\noindent{}where $r$ is the radial distance from the pore axis, $r_0$ = 6~\AA{}, $\sigma_r$ = 1~\AA{} and similarly for the axial coordinate $z$ with $\pm{}z_0$ = 19~\AA{} and $\sigma_z$ = 1~\AA{}.


All three maps were recorded on three-dimensional grids with spacing of 1~\AA.
In the remainder, any small discrepancies between volumetric map values computed in VMD and NAMD due to different interpolation methods (linear vs.\ cubic splines) are ignored; for additional details on this issue, consult the Multi-Map tutorial.\cite{multi-map-tutorial}

The maps were loaded into VMD, and each was used to define a CV based on the \texttt{mapTotal} Colvars function.\cite{Fiorin2020}
The number of water molecules in the cylinder, as well as the values of the three variables computed by Colvars/VMD are plotted in Fig.~\ref{fig:map-cv-eq}D-F, along with their block-averages over 1~ns and 10~ns segments.

\begin{figure}[h]
  \centering
  \includegraphics[width=\textwidth]{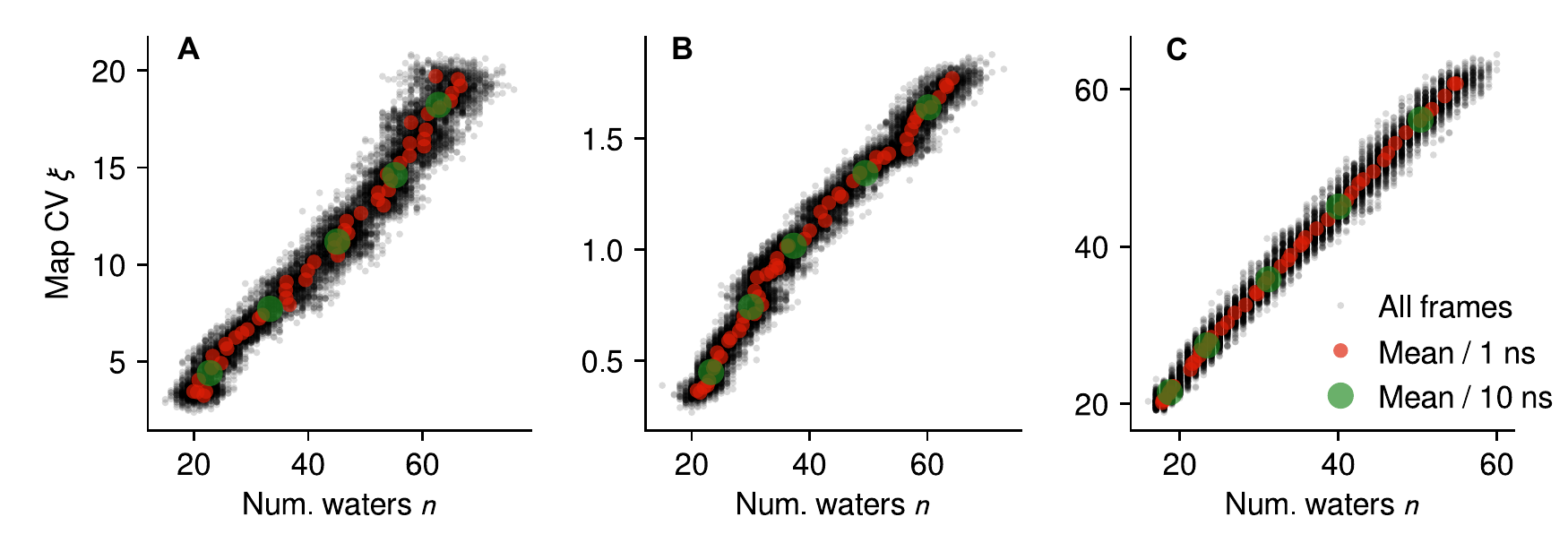}
  \caption{Values of volumetric-map CVs during 50~ns-long steered-MD simulations, with the restraint center initialized near the initial value of each variable and directed toward zero.  Similarly to \protect{Fig.~\ref{fig:map-cv-eq}}, the CV defined from the occupancy histogram \textbf{(A)}, the smoothed density map \textbf{(B)} and the cylindrical step function \textbf{(C)} are shown.}
  \label{fig:map-cv-smd}
\end{figure}

Steered-MD simulations were carried out using a moving harmonic restraint with a force constant of approximately $(1,000~\mathrm{kcal/mol})/{\langle\xi\rangle_{\mathrm{eq}}}^2$, where $\langle\xi\rangle_{\mathrm{eq}}$ is the mean value of $\xi$ in each equilibrium MD simulation (Fig.~\ref{fig:map-cv-eq}D-F); the center of the restraint was initialized at $\xi_0 = \langle\xi\rangle_{\mathrm{eq}}$ and moved toward $\xi_0 = 0$ in 50~ns.
The trajectories of the values of $\xi$ (Fig.~\ref{fig:map-cv-smd}A-C) show that all three definitions are suitable for artificially changing the number of water molecules in the transmembrane pore, with the corresponding water molecule count $n$ decrease from $\sim$60 to about $\sim$20.
The latter value is approximately the number of waters at the two interfaces between the pore and the outer water buffer, indicating that full dewetting of the transmembrane pore was achieved in all three simulations.

Nonetheless, the degree to which each map CV $\xi$ may be used as a proxy for the exact number of waters varies with respect to how smooth the map is.
This is particularly true when considering the relatively small equilibrium fluctuations (Fig.~\ref{fig:map-cv-eq}D-F).
For example, using the exact histogram (\texttt{volmap occupancy}) leads to relatively poor correlation between the map CV, $\xi$, and the number of waters, $n$ (Fig.~\ref{fig:map-cv-eq}D).
Using the smoothed density map (\texttt{volmap density}) largely improves correlation, both at the equilibrium distribution (Fig.~\ref{fig:map-cv-eq}E) and under the moving restraint (Fig.~\ref{fig:map-cv-smd}B).
Lastly, the most accurate match between $\xi$ and $n$ is achieved with a cylindrical map, which implements the same selection criteria making up for $n$ following the same cylindrical shape.

The above behavior is attributable to the spatial location of the gradients of each map, which in turn determine the atomic forces.
As long as regions with lower water density (e.g.\ at the pore's boundaries) are geometrically accessible to water molecules, forces aimed at decreasing the corresponding $\xi$ may lead to movement of water molecules along the membrane plane.
On the other hand, if the map decays to zero but far outside of the pore's boundaries (e.g.\ near the lateral face of the cylinder), forces are only applied to atoms positioned near the two interfaces with the outer buffer (e.g.\ the two end faces of the cylinder), and only the latter achieves transport inside or outside of the pore.

This example illustrates two important facts:
\begin{enumerate}

\item The Colvars Dashboard is useful at speeding up the process of defining a suitable CV based on volumetric maps.
  In the example shown, best agreement between the CV $\xi$ and the size of the VMD selection $n$ were achieved by using a smoothed version of the same step function that defines the selection itself.
  Alternatively, the next best choice was provided by a smoothed density map computed from equilibrium MD.

\item Through analysis of equilibrium and non-equilibrium trajectories using the Colvars Dashboard, it becomes easier to define statistically the relationship between the CV $\xi$ and another quantity $n$, which is in turn more intuitive to understand but at the same time cannot be used as a CV because it is intrinsically discontinuous.
  That same relationship may be used later to express PMFs computed along $\xi$ as a function of the second more intuitive parameter.
  This is particularly useful when dealing with \emph{Multi-Map} variables, where the mean values of a relatively complex CV $\xi$ can be expressed in terms of a more physically intuitive parameter.\cite{Fiorin2020}
  
\end{enumerate}